# Security Policy Specification Using a Graphical Approach

James A. Hoagland   Raju Pandey   Karl N. Levitt
**Department of Computer Science**
**University of California, Davis**
{hoagland,pandey,levitt}@cs.ucdavis.edu

A security policy states the acceptable actions of an information system, as the actions bear on security. There is a pressing need for organizations to declare their security policies, even informal statements would be better than the current practice. But, formal policy statements are preferable to support (1) reasoning about policies, e.g., for consistency and completeness, (2) automated enforcement of the policy, e.g., using wrappers around legacy systems or after the fact with an intrusion detection system, and (3) other formal manipulation of policies, e.g., the composition of policies. We present LaSCO, the Language for Security Constraints on Objects, in which a policy consists of two parts: the domain (assumptions about the system) and the requirement (what is allowed assuming the domain is satisfied). Thus policies defined in LaSCO have the appearance of conditional access control statements. LaSCO policies are specified as expressions in logic and as directed graphs, giving a visual view of policy. LaSCO has a simple semantics in first order logic (which we provide), thus permitting policies we write, even for complex policies, to be very perspicuous. LaSCO has syntax to express many of the situations we have found to be useful on policies or, more interesting, the composition of policies. LaSCO has an object-oriented structure, permitting it to be useful to describe policies on the objects and methods of an application written in an object-oriented language, in addition to the traditional policies on operating system objects. A LaSCO specification can be automatically translated into executable code that checks an invocation of a program with respect to a policy. The implementation of LaSCO is in Java, and generates wrappers to check Java programs with respect to a policy.

## 1.0  Introduction

Security forms a core component of any system. What is meant by security has been described in general terms such as confidentiality, integrity, and availability. However, what is precisely meant by security (or for something to be secure) for a particular system[1] varies from system to system and possibly depending on the situation. The military would likely have a different definition of what the appropriate security is for their system than a bank, a university, or a home user would. They each have their own needs that should be reflected when securing their systems. In addition, one can describe security at various system levels. A security policy is a description of the security goals for a system and how a system should behave in order to meet these goals. An example policy is that the request for a purchase and its approval must be from different users.

Since the security goals for a system effect how the security mechanisms on the system are configured, it is important for the security policy to be stated clearly. Additional benefit is obtained if the policy can be directly used to configure the security mechanisms or can be used to formally reason about the effect of the policy. A common security need is to restrict access to the resources on a system. This is reflected in a constraint security policy, which states constraints on the system[2]. We focus on constraint security policies in this paper.

---

1. We use "system" generally here. It can be almost anything on a computer that contains some sort of entities and can be interacted with or can be seen as executing. Some examples are: a program executing, a file system, an operating system, a workgroup, and a network of hosts.
2. Other types of security policies are ones that describe how trust is managed, describe how to choose the security options in setting up a network connection, and ones that describe how to respond to a security violation.

It is inadequate to formulate policy just in terms of operating system objects such as files and network connections. For some situations, we need to state policy at a level beyond which an operating system has good knowledge, for example at the application level. To an operating system, an object in an application is typically not distinct and the operating system would in any case have less knowledge of the semantics of an object - what it means and represents. Thus, it is important for a means to state policy to be able to operate at this level as well as at lower levels.

Our view of policy is object-based. Typically a system would contain one or more objects. These objects do not exist in isolation. They interact through events such as accesses and creation methods. Policy puts constraints on the events that interrelate the objects and on the objects themselves. A particular policy is in effect over some certain domain. This domain describes when the policy is relevant. This might reflect the state of an object, that a particular event has occurred, or that some set of events has occurred. Once we have a place where the policy is relevant, the requirement of the policy becomes important. The requirement makes a statement of how the system should be. It might place a restriction on the particulars of an event such as requiring a certain type of event to be used, place a restriction on the state of an object, or some combination of these.

In this paper, we define a language that we call LaSCO (the Language for Security Constraints on Objects) which represents policy as a directed graph with annotations. We use this form this experimentally for reasons of visual convenience. There are several reasons we find LaSCO appealing. LaSCO can be used to state policies for a variety of types of systems. Notably we have found that it can state policies on operating systems and polices for a programming language, particularly if it is object oriented. The policy can be specified separately from a program. The language is not tied to any particular enforcement mechanism and the policies stated in it may be used in different ways. For example, policies may be compiled into a program, used by a reference monitor [1] to mediate access to system objects, or checked in an off-line mode against an audit trail.

We can state policies that apply whenever a certain pattern of accesses is encountered. This can be seen as a template for a general policy rather than the set of configuration details that an access control matrix describes. The language can describe policies that span several events over a period of time and can describe policies that depend on conditions other than the state at the time of check.

The language is visual which might make it easier for humans to deal with policies stated in the language. At the same time, the language has a formal basis. Having a formal basis and a formal semantics promotes our ability to reason about policies and about system with policies employed.

This paper is organized as follows: Section 2.0 presents an overview of our approach. In Section 3.0 we present our model of the system to which a LaSCO policy is applied and in Section 4.0 we present the language and its semantics. We present several examples of LaSCO policies in Section 5.0. Section 6.0 describes some operations on LaSCO policies. Section 7.0 addresses both the expressiveness and the limitations of the language and present some extensions to LaSCO to address this. Our plans for implementation are presented in Section 8.0 and Section 9.0 presents a comparison of our work with related work. Section 10.0 contains a summary of our approach and discusses future work.

## 2.0 Overview

We describe security policies using a formal language based on directed graphs. The system to which the policy is applied consists of a series of events, each occurring between a pair of objects. We apply the policy to the system by identifying portions of the system to which the policy applies and checking the requirement part of the policy on that portion of the system.

In Section 2.1 we give an overview of our model of a system to which policy is applied. Section 2.2 gives an overview of our policy language. We then present two example applications of the language. Section 2.3 demonstrates a way to apply polices to an object oriented programs and we describe applying policies to a file system in Section 2.4.

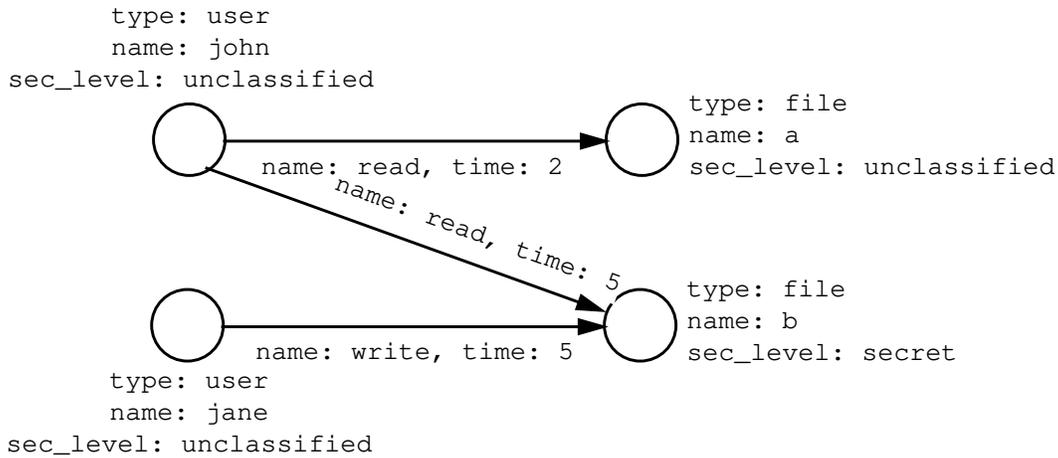

**Figure 1.** Example system graph

## 2.1 System model overview

A system instance is a snapshot of the system at a moment in time; it contains a set of active objects and a set of pending events. Events are interactions between two objects. Pending events are ones that are intended to execute, but which are awaiting approval. It is at this point where policy is considered for a system. The result of applying a policy to the system is knowledge of whether and how the system violates policy; the entity doing this checking may use this knowledge however it deems appropriate. We assume that events occur at discrete time steps and that multiple events can be happening simultaneously. We associate the time of a system instance with each event in it using a *time* parameter.

A system consists of a sequence of system instances. At a particular time, an object has a constant set of attribute values associated with it. However, the object has different sets of attribute values at different times, but a fixed set of attribute names. Each event has a set of parameter values associated with it, as well as a source and destination object.

A system can be precisely represented as an annotated graph that we term a ***system graph***. There is a node for each object and an edge for each event. The edge is between the source of the event and the target of the event. The set of attribute sets on an object is an annotation of each node and the set of event parameters is a annotation on each edge.

We present an example system graph (representing a certain system) in Figure 1. There are 4 objects and 3 events depicted there. If an object in the system had changed its attribute values, then more sets of attributes would also be associated with it and would also be part of the node annotation for the corresponding node.

## 2.2 Language overview

The purpose of the policy language is to describe constraints on a system that must hold in certain situations that occur in the system. When applied to a system, these situations might span several events, involve a single event, or just involve objects. The requirement can involve restrictions on object attribute values, event parameter values, and relationships between these. We use a policy graph to represent both the situation under which a policy applies and its requirement. A LaSCO policy graph can be partitioned into two parts: the *domain* and the *requirement*. The domain describes the situation under which the access control policy should be in effect and the requirement describes the constraint that must hold for the policy to be upheld. A security policy is, thus, stated by specifying that if a system is in a specific state (as specified by the domain), a specific access constraint (specified by the requirement) must hold for that system.

To describe the domain and requirement, we extend directed graphs with annotations. These annotations are termed predicates and there is one for the domain and one for the requirement for each node and edge. The function of these logical conditions is to describe the actual objects or events that we intend the node or edge to represent. We sometimes view this as trying to match a predicate to an object or event on the system. From this point of view, we evaluate the predicate in the context of a particular object's attributes or a particular event's parameters. The nodes and edges in the policy graph along with their domain predicates form the representation of the domain of the policy. The requirement predicates on nodes and edges form the requirement of the policy. Effectively describing policies using LaSCO is enabled by support for matching attribute values, arbitrary predicates, and Prolog-style logical variables.

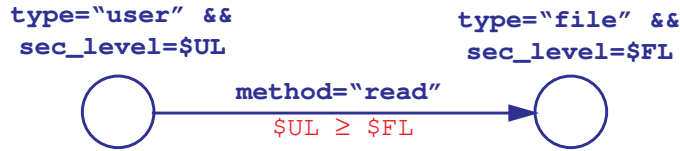

**Figure 2.** Policy graph for the simple security property

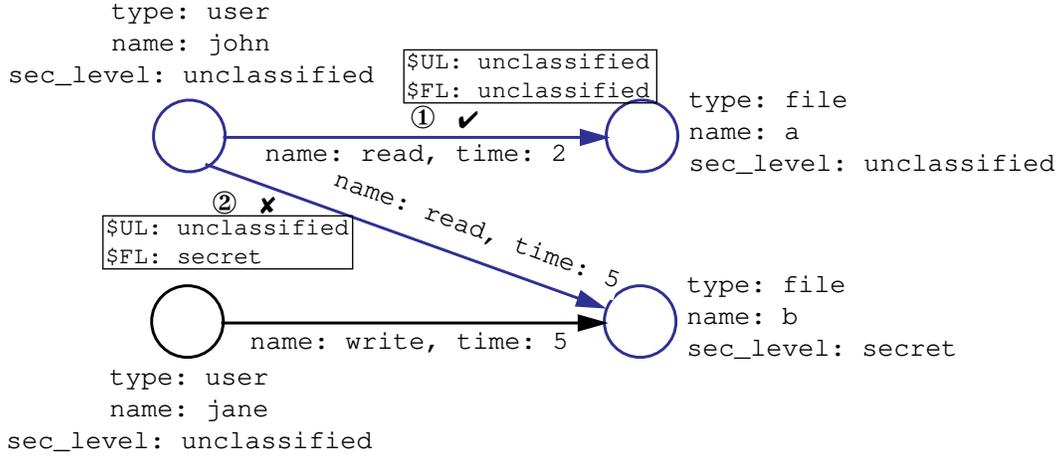

**Figure 3.** Depiction of simple security property applied to the example system graph. The two places where the domain applies is noted with the necessary variable bindings.

We apply policy to a system by finding locations in the system graph where the nodes and edges of the policy graph match, given that the domain predicates match the respective objects and events. For each of these matched locations, we test the requirement of the policy by evaluating the requirement predicates for each policy graph node and edge. This is done in the context of the system object or event that node or edge matched for the domain. If each of the requirement predicates evaluate to true, then the policy is upheld, otherwise the policy had been violated by the system.

For purposes of illustration, we now present a simple LaSCO policy graph. The LaSCO representation of the simple security policy of Bell-LaPadula [2] is depicted in Figure 2. The policy specifies that if a user is reading a file, the security level of the user must be at least as great at that on the file.

In policy graph depictions in this paper, the **blue bold text** near a node or edge is the domain predicate of that node or edge, and the red standard text annotation is the requirement predicate. Nodes and edges without an explicit predicate for either the domain or the predicate have an implicit "True" expression as that predicate.

In this graph, the left node denotes a user object, whereas the right node denotes a file object. The edge between the nodes denotes an access relationship as specified by a read event. The access control graph has two components. The domain specifies that the policy holds when the access relationship "read" occurs between a user and a file. Type constraints are used to select specific objects. The requirement specifies that the level of user should be at least as high as that of the file. This is specified by annotating the graph with the expression $UL ≥ $FL. Note that $UL and $FL denote logical variables that get bound to sec_level attributes of the user and file objects respectively.

In Figure 1, we consider the example system depicted in figure 1 in the context of the simple security property. When the domain is applied to the system, the policy graph matches in two locations. One is with the user node matching the object with "name=john" attribute, the file node with the "name=a" object, the edge with the top read event, and with $FL bound to "unclassified" and $UL bound to "unclassified". "$UL ≥ $FL" evaluates to true so the requirement is satisfied for this location.

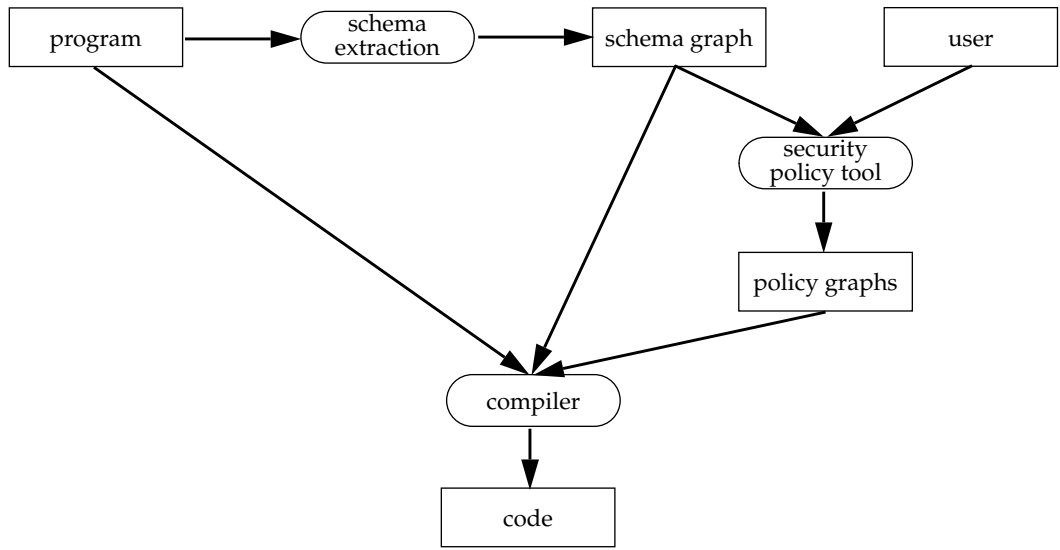

**Figure 4.** Access specification and control for applications at the language level.

The second place that the policy matches is with the user pattern node matching the "name=john" object, the file pattern node with the "name=b" object, the pattern edge with the lower read event, and with $FL bound to "secret" and $UL to "unclassified". In this case, the requirement is not satisfied since the requirement predicate is not satisfied with $UL at "unclassified" and $FL as "secret", so the policy is violated. The entity that wished for the policy to be tested for the system may wish to block that event. Where there is a write event between the "name=jane" user object and the "name=b" file object, the domain pattern nodes match, but the pattern edge does not, so this particular policy does not apply.

## 2.3 Using LaSCO to enforce policy on a program

We view a program with associated security policies as a tuple ‹D, P›, where D denotes a definition of the program and P denotes a set of policies that make the program meet the users security goals (at least as stated in the policy). We assume that an object-oriented programming language is used for defining programs. Thus, D denotes a collection of class specifications. Each class defines a set of instance variables (denoting states of the instances of the class) and a set of methods. The execution of the program is the system for the policies. In the system, objects are instantiated and methods are invoked to perform computations. We note that access relationships among objects and methods occur when objects are instantiated and methods are invoked. Access control, thus, involves establishing constraints on object instantiations and method invocations. LaSCO facilitates definition of such access constraints. For this type of system, program objects are the objects for the system model and method invocations are the events.

Note that most traditional programming languages do specify mechanisms (such as types and public/private methods and variables) for controlling accesses to object states and methods. However, most such constraints are static. They do not allow specifications of any access control that is dynamic and that depend on the state of an application. Note that while it may be possible to add constraint checks directly in a program, for example, while writing the program, there is added benefit from stating the policies separately using a formal method. These include allowing some degree of independence between the program and the policy, facilitating better understanding of the effect and intention of a constraint through a directed language, and enabling reasoning about policies and interactions, independently of a particular system.

We now describe the manner in which access control policies might be derived, specified and enforced. In Figure 4, we show a typical set of steps taken by a user for specifying access control policies:

- Schema Extraction: Given an arbitrary program, a schema extraction tool constructs a program schema graph from the program. Nodes of the program schema graph denote class definitions of the program, whereas edges between the nodes represent method invocations or object instantiations found in the program source. Edges, thus, capture the various access relationships. The program schema graph can be presented to the user.

- Security policy tool: The user selects a portion of the program schema graph using a security policy tool. She then constructs a LaSCO policy graph by adding constraints over the schema graph. The constraints specify the condition under which accesses are permitted. In addition, the security policy tool will provide a library of security policy graphs that the user can customize for the schema. This leads to the creation of a set of policy graphs.

- Compiler: A compiler takes the program, access control constraints represented in policy graphs, and a program schema graph and generates code for both implementing the program and for enforcing the access constraints. An important component of our research is to develop techniques which will make the process of adding enforcement code incremental. Thus new policies can be added or existing policies modified without requiring recompilation of the whole program.

### 2.4 Using LaSCO for file systems

We can describe access control policies for a file system using LaSCO. To do this, we might model the file system as consisting of files and subjects (for example, users or processes), both of which would be objects in the LaSCO system model. Files objects would naturally have attributes such as its owner, type, size and modification date, and the subject objects would have whatever security relevant attributes are appropriate for the system. The system events would include accesses by subjects to the files on the system with the details of the access as parameters. Policies for this system thus might restrict access to files based on the details of the access, what is being accessed, and what has previously occurred on the system. Policies also might restrict the state of files.

The policy engine enforcing the LaSCO policies might serve as a reference monitor, moderating requests to the file system from applications. It would monitor access requests and attribute changes and block any action that would violate the given policies. To do this, for some policies, it would need to maintain a history of accesses, possibly in a summary form. An alternate way in which a policy enforcement engine might be situated is having the file system make calls to the engine when a policy decision is needed. The engine would then return whether or not the action is permissible, and the file system could act on that as it sees fit. Thus it could sit alongside the file system, but be fairly tightly coupled. A third possibility would have the access checking done in an off-line manner by an application such as an intrusion detection system that would scan over an audit log of accesses and object states. Though this precludes the possibility of preventing unauthorized access when detected, this might be a good alternative in some situations, such as when the use of LaSCO policies is being retrofitted onto a system.

### 3.0 The System Basis for LaSCO

Policies are in effect over some system. We model the system upon which a policy is applied so that we have a construct upon which the definition of LaSCO is based. With a system model, we can define the meaning of LaSCO as it applies to an arbitrary system that fits the system model.

### 3.1 LaSCO system model

The LaSCO system model is object-based. It consists of some set of objects and events that occur between the objects. Events have unchanging named parameters associated with them. Objects consist of a fixed set of named attributes which describe the security-relevant state of the object. The values of these attributes may vary over time. However there is an unchanging and unique 'id' attribute on each object.

We view a system as consisting of a series of system instances, each representing the system at a particular time. In a system instance, the objects present are those that are active and the events are those wishing to execute. In an instance, an object's attribute values are fixed. We formally define a system and system instance here:

**Definition 1.** System instance.
  Let ⟨$M_t,O_t$⟩ denote the system instance at time t, where:
    $M_t$ is a set of pending events and
    $O_t$ is a set of currently active objects.

**Definition 2.** System.
  Let ⟨I,<,VALUES⟩ denote a system, where:
    N is the set of natural numbers
    I={⟨$M_t,O_t$⟩ | t∈ N} is the set of all system instances in a system,
    < is a relation that totally orders system instances in I by the time they occur (⟨$M_i,O_i$⟩ < ⟨$M_j,O_j$⟩ iff i < j), and
    VALUES is the set of all possible values of attributes and event parameters in the system.

  Now we define some notation for objects and events here.

**Definition 3.** Object attribute values.
  For each attribute of an object, the object's set of attribute values contains a tuple of the form ⟨α,μ⟩ where:
    α is the attribute name, and
    μ is the value of α

**Definition 4.** System object.
  Let o denote an object in a system instance:
    o is the set of attributes associated with the object, each as in Definition 3.
    id(o)=i where ⟨'id',i⟩ ∈ o

**Definition 5.** Event parameter values.
  For each parameter of an event, the event's set of values contains a tuple of the form ⟨ρ,μ⟩ where:
    ρ is the parameter name, and
    μ is the value of ρ

**Definition 6.** System event.
  Let ⟨s,d,p⟩ denote an event for a system instance, where:
    s is the id of the source object of the event,
    d is the id of the destination object of the event, and
    p is the set of event parameters associated with the event, each as in Definition 5.
  For event e=⟨s,d,p⟩:
    src(e)=s
    dest(e)=d
    param(e)=p

Note the similarity between object attributes and event parameters.

## 3.2 System Graph

A *system instance graph* is a way to represent a system instance as a graph. It contains a node for each object in the instance and an edge for each pending event, from the node corresponding to its source object and to the node corresponding to its destination object. The edges are annotated with event parameter bindings for the corresponding event. The attribute bindings of objects in the instance are annotations on the corresponding node.

One can view a set of system instances as a graph as well. In a *system graph*, the nodes and edges are all the objects and events in the set. The event parameters are annotations on edges as in the system instance graph. In addition, a 'time' parameter denotes the particular system instance in which the event occurred. System graph nodes contain a set of sets of attribute values. Each set of attribute values is the state of the object at a particular system instance. There is such a set at a node for each system instance graph in which the node is incident with an edge. This is essentially the overlaying of the system instance graphs for all system instances. We define some system graph notation in Definition 7.

**Definition 7.** System graph.
  Let ⟨O,M⟩ denote a system graph for the system instances of the times in T, where:

O={o | t∈T ∧ o ∈ $O_t$} is the set of nodes of the graph, and

M={‹s,d,(p ∪ {‹'time',t›})› | t∈T ∧ ‹s,d,p› ∈ $M_t$} is the set of edges of the graph.

For each system graph edge e∈M:

time(e)= t where ‹'time',t› ∈ param(e) is the time the event corresponding to the edge occurred

src_attr(e)=o where o ∈ O ∧ time(o)=time(e) ∧ src(e)=id(o) is the set of attribute values on src(e) that is from the same system instance as e

dest_attr(e)=o where o ∈ O ∧ time(o)=time(e) ∧ dest(e)=id(o) is the set of attribute values on dest(e) that is from the same system instance as e

When we apply a policy to a system, we match the graph of a policy against the system graph that represents the series of system instances up to a certain time (the "current" time) in the system.

## 4.0 Access Control Policy Language

### 4.1 Graphs

For LaSCO and for defining its semantics, we use several kinds of graphs. We defined a system instance graph and system graph in Section 3.2. We use these to represent one or more system instances. To keep our terminology clear, we will refer to a conventional directed graph as a ***basic graph***, which consists of basic nodes and basic edges. Definition 8 gives the notation associated with basic edges and Definition 9 that with basic graphs.

**Definition 8.** Basic edge.

Let the tuple ‹s,d› denote a basic edge, where:

s is the source node and

d is the destination node

src(‹s,d›)=s

dest(‹s,d›)=d

For a node n, isolated(n)=‹∀e : e ∈ E : src(e) ≠ n ∧ dest(e) ≠ n› is whether n is a isolated node

**Definition 9.** Basic graph.

Let the tuple ‹N,E› denote a basic directed graph, where:

N is the set of basic nodes in the graph, and

E is the set of basic edges in the graph.

A *isolated* node is a node that has no incident edges.

Policies in LaSCO are represented in a policy graph. The two parts of the policy, the domain and the requirement, may be thought of as a pattern graph. We elaborate on the definition of policy graph and pattern graph in the next section.

### 4.2 Language description

We use policies described in LaSCO to find matches in the system (or, equivalently, the system graph) for where the policy applies. Then we check the requirement of the policy against this match. Thus, the policy serves as patterns on the system.

In LaSCO these patterns are represented by a ***policy graph.*** A policy graph is an annotated directed graph which contains two parts: the ***domain*** and the ***requirement***. The domain is a pattern for the system which describes when policy is in effect. The requirement, on the other hand, is a pattern which indicates the restrictions imposed on the system by the policy. Both the domain and requirement consist of nodes and edges annotated with predicates. We call the predicates associated with domain and requirement ***domain predicates*** and ***requirement predicates*** respectively. For instance, there are three domain predicates (`type="user" && sec_level=$UL`, `method="read"`, and `type="file" && sec_level=$FL`) and one explicit requirement predicate ($UL ≥ $FL) in Figure 2[3]. LaSCO policies make use of a set of a policy variables to relate different attribute values and parameter values of different objects and events. The scope of variables is a single LaSCO policy graph.

---

3. If a domain or requirement predicate is not explicitly stated for a node or edge, its implicit value is "True".

Definition 10 defines the notation for a policy graph:

**Definition 10.** Policy graph.
Let a policy graph be represented by the tuple ‹G,γ,λ,V›, where:
G is the basic graph in the policy graph,
γ is a function mapping elements of G to their domain predicate,
λ is a function mapping elements of G to their requirement predicate, and
V is the set of variables for the policy graph.

We find it convenient to view the domain and requirement each as a ***pattern graph***, a pattern for part of a system. Definition 11 gives the notation associated with pattern graphs.

**Definition 11.** Pattern graph.
Let a pattern graph be represented by the tuple ‹G,χ,V›, where:
G is a basic graph,
χ is a function mapping each node and edge in G to its predicate, and
V is a set of policy variables associated with the pattern graph.

A node n in G along with its predicate χ(n) are termed a ***pattern node***. A pattern node represents an entity in our policy that matches a set of objects in a system. For example, the left domain pattern node in Figure 2 represents all users and would match any object with "type" attribute equal to "user". A ***pattern edge*** is an edge in G with its predicate. A pattern edge matches a set of events. The one domain pattern edge in Figure 2 represents a read event from a user object to a file object.

The domain pattern graph consists of a basic graph, the domain predicates, and the set of variables mentioned in the policy. This is a pattern to apply to the system which matches when the policy applies. The requirement pattern graph consists of a basic graph, the requirement predicates, and the set of variables mentioned in the policy. This matches when the policy is upheld by a portion of the system. These two pattern graphs are related in that they have the same basic graph and the same set of variables are used in both graphs. (Since we have the same basic graph, we can depict the graphs superimposed.) The domain for a policy is ‹G,γ,V› and the requirement is ‹G,λ,V›. Note that together these two pattern graphs fully describe a the policy graph ‹G,γ,λ,V› as defined in Definition 10.

***Variable bindings*** represent a set of policy variables that have values bound (assigned) to each of them. Definition 12 presents some notation associated with variable bindings.

**Definition 12.** Variable bindings.
For each variable in a variable binding B, B contains a tuple of the form ‹υ,μ› where:
υ is the variable name, and
μ is the value bound to υ.
vars(B)= { υ | ‹υ,μ› ∈ B} is the set of variable names in B
val(B,υ)= μ where ‹υ,μ› ∈ B is the value in B of υ

A set of variable bindings is said to be ***complete*** if every policy variable (in the current context) has a value bound to it. A pattern graph is satisfied when all pattern nodes and pattern edges can together be satisfied by a set of variable bindings. When a policy is being applied to a system, it is up to the mechanism doing this application to determine the variable bindings which allow this, if indeed there is such a set. This determination would be based on the predicates and the system objects and events.

Logically, a predicate is a pattern on the actual objects or events. This pattern is a statement as to which objects and events the pattern node or pattern edge represents. Predicates make references to the attributes of objects and parameters of events and (through variables) to attributes and parameters of other objects and events.

Syntactically, a predicate is a boolean expression containing the logical operators "&&" (and), "||" (or), and "!" (not); the (polymorphic) comparison operators "=" and "!=" (not equal); the numeric

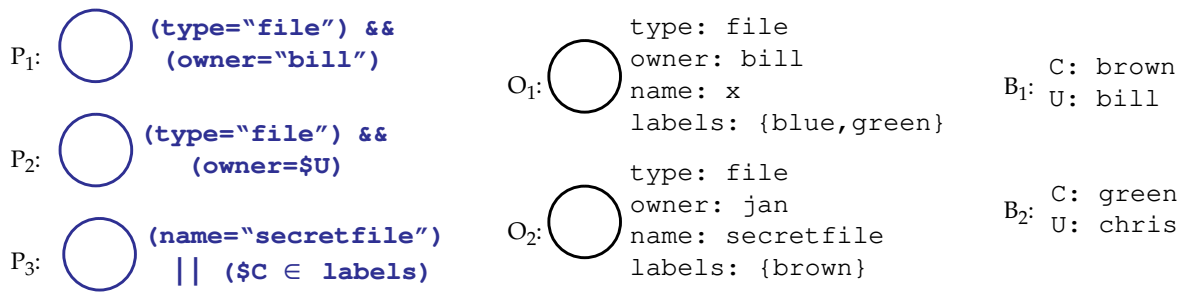

**Figure 5.** Example pattern nodes, system objects, and variable bindings. $P_1$, $P_2$, and $P_3$ are pattern nodes, $O_1$ and $O_2$ are system objects, and $B_1$ and $B_2$ are variable bindings.

|       | $O_1$ | $O_2$ |
|-------|-------|-------|
| $P_1$ | satisfied with any variable bindings | not satisfiable by any variable bindings |
| $P_2$ | satisfied by $B_1$ but not $B_2$ | not satisfied by $B_1$ or $B_2$ |
| $P_3$ | satisfied by $B_2$ but not $B_1$ | satisfied by either $B_1$ or $B_2$ |

**Figure 6.** Predicate evaluation example. The table depicts which variable bindings satisfy each pattern node's predicate given the object.

comparison operators "<", ">", "<=", and ">="; the set element test operator "∈"; the set comparison operators "⊂" and "⊆"; basic mathematical operations; set operations "∩" and "∪"; and arbitrary nesting of expressions through parentheses. Each predicate can name only attributes associated with an object on the system (when associated with a node), parameters of events (when associated with an edge), policy variables (denoted with a "$" prefix), and constants. A predicate is evaluated with respect to a particular object or event and a set of variable bindings. Attributes named in the predicate evaluate to the value of the attribute on the associated object, parameters to the event parameter value on the associated event, and variables to their bindings in the given variable bindings.

There are two restrictions on what may and may not be found in predicates, given its context. Each variable present in the policy must be in some subexpression of some domain predicate as <variable>=<value> (or <value>=<variable>) where <value> is a (possibly derived) single value. This subexpression may not be part of a disjunction. This ensures that all variables have a single value for the domain. The second restriction is that node requirement predicates may not contain any (direct) attribute references[4].

Figure 5 presents several example pattern nodes, system objects, and variable bindings. The table in Figure 6 uses these to give examples of evaluating a predicate on a domain pattern node in the context of a system object and variable bindings.

A predicate in the requirement is used differently than one in the domain, although both are syntactically the same and evaluated with respect to an object's attributes or an event's parameters and a set of variable bindings. A requirement predicate is a necessary condition for the requirement of the policy to be met. Unless every requirement predicate evaluates to true given the same variable bindings as in the domain, the policy is violated.

---

4. The values of an attribute might change during the events that are incident to the object. This would lead to ambiguity if the attribute is mentioned in a requirement predicate. Thus, we impose this restriction to alleviate this ambiguity. Variables may be bound to an attribute value in the domain and referred to in the requirement, which ensures that the significant attributes only match the policy in one particular way while holding a particular value. This restriction could be narrowed to nodes that have more than one incident edges without ambiguity arising.

## 4.3 Semantics of LaSCO

### 4.3.1 Matching predicates

#### 4.3.1.1 Variable conditions

For the purposes of formalizing LaSCO's semantics, we introduce the concept of *variable conditions*, the logical conditions for variables under which a predicate is satisfied. Variable conditions consist of variable bindings and a *condition expression.*

**Definition 13.** Variable conditions.
Let variable conditions c be represented by the tuple ‹$B_c,C_c$›, where:

$B_c$ is the (possibly not complete) set of variable bindings for c as defined in Definition 12, and

$C_c$ is the condition expression for c.

A variable is bound and is present in the variable bindings only if it can be satisfied by a single value. A condition expression indicates the possible values for those variables that are not bound to a single value. This expression, when given particular values to use for the currently unbound variables, will evaluate to true if the new bindings are acceptable and false otherwise. The condition expression may contain anything that a predicate may contain except for attribute references and parameter value references. Example condition expressions are "$\$x > 17$ && $\$y \in \{a,b\}$", which indicates that the variable x must have a value greater than 17 and that the variable y must have either a or b as its value; and "$\$x > \$y$", which indicates that the variable x must have a value greater than the variable y for the context of the variable condition to be satisfied.

#### 4.3.1.2 Predicate evaluation

When we evaluate predicates, we do so in the context of a set of attribute values or event parameters and with respect to a set of variable bindings. Predicates are evaluated with respect to variable bindings and either a system object or event. When evaluating a predicate, we simultaneously substitute names in the predicate for these values.

**Definition 14.** Substitution for values.
We denote substitution of a set of names for their values by $p \bullet \theta$, where:

$\theta$ is either a set of variable bindings (as defined in Definition 12), a set of attribute values (as defined in Definition 3), or a set of event parameters (as defined in Definition 5), and

p is a predicate or condition expression

In evaluating a predicate expression, we substitute the value of a variable for the name of the variable and substitute the value of an attribute or parameter for its name. If the variable bindings are complete, what results from this is a constant expression which evaluates to true or false. Formally stated, given a predicate p, a set of attribute values or event parameter values $\sigma$, and a set of variable bindings B, we define sat_pred(p,$\sigma$,B) to be the variable conditions under which p evaluates to true given $\sigma$ and B:

**Definition 15.** sat_pred.
sat_pred(p,$\sigma$,B)= ‹B,(p$\bullet\sigma$)$\bullet$B›

This is obtained by forming a variable condition tuple with B and the condition expression resulting from substituting $\sigma$ and B in p.[5]

#### 4.3.1.3 Merging variable conditions

A context is a set of predicates evaluated under the same variable bindings and having a mutual set of variable conditions under which each of the predicates holds. Predicates are sometimes applied on different objects or events simultaneously in a mutual context with their mutual policy variables each pos-

---

5. In the case that an attribute or parameter name is mentioned in a predicate but is not found in s, the most immediate boolean expression in which the name evaluates to false, regardless of any other part of that expression. This implies that, unless that boolean expression is within an expression with a disjunction, this forces the predicate not to be satisfied by the object or event.

---

sessing a value acceptable by all predicates. We need this for matching pattern graphs to the system as described in Section 4.3.3 below.

Thus, we define a function, merge_conds, that takes a number of variable conditions from different contexts and merges them into the variable conditions for a larger combined context. If there are variable bindings in the conditions that are not equal, then the merged variable condition becomes ‹{},false› (meaning there is no way to satisfy all contexts), otherwise it becomes the reduced version of the condition formed by taking the union of the variable bindings and the "and" of the condition expressions. Reducing a condition is finding all bound variables in a condition expression and moving them to the variable bindings.

**Definition 16.** merge_conds.
merge_conds(c1,c2, …, cn)= merge_conds(merge_conds(c1,c2), …, cn)
merge_conds(c1,c2)={‹{},false›     if $‹\exists \upsilon : \upsilon \in (vars(B_{c1}) \cap vars(B_{c2})) : val(B_{c1},\upsilon) \neq val(B_{c2},\upsilon)›$
        {reduce_cond($‹B_{c1} \cup B_{c2}, C_{c1} \wedge C_{c2}›$)     otherwise
reduce_cond($‹B_c,C_c›$)= $‹B_c \cup B, C›$ where r=$‹B_c,C_c \bullet B_c›$ and ‹B,C›=extract_bound(r)
extract_bound(c)= ‹B,C›, where B is the set of variable bindings found in $C_c$ and C is $C_c$ with the bindings removed

### 4.3.2 Pattern node and pattern edge matching

A pattern node is a template for matching system objects. The predicate associated with a pattern node determines the type of system objects that match it. If the predicate evaluates to true given an object and some set of variable bindings, then the pattern node matches the object for the variable bindings.

The variable conditions under which a pattern node with the predicate p matches the system node attributes σ with variable bindings B are given by match_node(p,σ,B). These are the conditions under which the predicate is satisfied by the attributes and the variable bindings.

**Definition 17.** match_node.
match_node(p,σ,B̄)= sat_pred(p,σ,B)

A pattern edge has a pattern node as its source that represents the originator of the event that the edge represents and a pattern node as its destination that represents the target object of the event. The variable conditions under which a pattern edge with predicate p matches the system edge parameters σ with variable bindings B are given by match_edge(p,σ,B). This is identical to the conditions under which the predicate is satisfied by the system edge parameter values and the variable bindings.

**Definition 18.** match_edge.
match_edge(p,σ,B̄)= sat_pred(p,σ,B)

A pattern edge together with its nodes matches an event using variable bindings B if:

1) using B, the predicate of the edge evaluates to true for the event,
2) the source pattern node matches the source object of the event using B and the attribute values that were on the source object at the time of the event, and
3) the destination pattern node matches the destination object of the event using B and the attribute values that were on the destination object at the time of the event.

### 4.3.3 Pattern and policy graph matching

Given a pattern graph ‹G,χ,V›, a mapping of edges in G to edges in a system graph ϖ, a mapping of isolated nodes in G to nodes in a system graph ω, and a complete set of variable bindings B, match_graph(G,ϖ,ω,χ,B) returns true if and only if the pattern graph matches given the pattern graph to system graph mappings and the variable bindings. This is true if, for each edge in G, the edge and its incident nodes match the corresponding edge and nodes in the appropriate system instance graph and for each isolated node in G, it matches the corresponding object. Let c denote the merged variable conditions resulting from matching the edge and its incident nodes given B. The edge and incident nodes match if $C_c$ is true[6].

**Definition 19.** match_graph.
   match_graph(G,$\varpi$,$\omega$,$\chi$,B)= ‹∀e : e ∈ E : $C_c$ where c=merge_conds(match_edge($\chi$(e),param($\varpi$(e)),B),
      match_node($\chi$(src(e)),src_attr($\varpi$(e)),B),
      match_node($\chi$(dest(e)),dest_attr($\varpi$(e)),B))›
   ∧ ‹∀n : n ∈ N ∧ isolated(n): $C_c$ where c=match_node($\chi$(n),$\omega$(n),B)›

Note that the order of evaluation of node and edge predicates in a pattern graph is irrelevant in LaSCO, that they are considered to be simultaneously evaluated. Note that if a particular predicate does not refer to attribute values or event parameters, it does not matter where it appears in the pattern graph.

### 4.3.4 Domain Matching

The domain is a pattern to apply to the system which matches wherever the policy is relevant. When we apply the domain of a policy successfully to a part of the system, it produces what we term a ***policy to system match*** (***match*** for short). A match represents the location and the way in which the domain is satisfied. As the domain may apply in several ways in the system or not at all, applying the domain to the system produces a set of matches.

A policy to system match consists of a mapping between the domain and the system and a set of variable bindings. The variable bindings are the ones that permit the association of pattern edges to system graph edges. The mapping between the domain and the system takes the form of a one-to-one correspondence between pattern edges and system events (or, equivalently, the system graph edges related to the pattern edges) and between isolated pattern nodes and system objects (with the attributes from a particular time). We will informally refer to the system events and system objects in this mapping as a *part of the system*. Note that at most one set of variable bindings will cause the domain to match a particular part of the system due to the requirement we mentioned in Section 4.2 that each variable be on one side of an "=" operator in some domain predicate. This has benefits toward implementation efficiency since once a set of variable bindings is found for a domain and part of a system, one can stop looking. A match completely describes a match of a policy domain to the system. Definition 20 formalizes the definition of a policy to system match:

**Definition 20.** Policy to system match.
   Let $\Theta_P$=‹$\varpi$,$\omega$,B› denote a complete policy to system match, where:
      P=‹G,$\gamma$,$\lambda$,V› is a policy graph as defined in Definition 10,
      $\varpi$ is a function mapping each edge in G to a system edge (E→M),
      $\omega$ is a function mapping each isolated node in G to a system node (N→O), and
      B is a complete set of variable bindings.
   We shorten $\Theta_P$ to $\Theta$ when the policy referred to is clear.

We require that the edge binding function $\varpi$ be one-to-one, so that each policy edge matches a distinct event.

The semantics of a node-match is that it must match for the attribute values for the system node for each system instance in which an incident edge matches. Since isolated nodes have no incident edges, they can match the system node in any system instance. An interesting question is what happens if the value of a object attribute changes such that it no longer matches a node as it did for a previous event that matched an adjacent edge. The semantics that the same set of variable bindings are used for all predicates and that nodes are matched when edges are matched answer that question. It does not matter if the attribute values change. The domain will match (for the part of the policy involving that node) when each of the edges find a match given the value of the object attributes at the time of the event that matches the edge.

The domain of a policy is satisfied if the domain pattern graph is satisfied by a match. We denote a satisfaction of the domain of policy graph P by $D_P$(‹$\varpi$,$\omega$,B›). Matches(P,S) is all the ways to satisfy $D_P$ with the system S. (Here $2^T$ denotes the power set of T.)

---

6. Since B is complete, the condition expression will not contain any variables and hence will evaluate to either true or false.

**Definition 21.** Domain satisfaction.
   $D_P(‹\varpi,\omega,B›)$ = match_graph$(G,\varpi,\gamma,B)$

   matches$(P,s)$ = $\{‹\varpi,\omega,B› \mid \varpi \in 2^{E \to M} \wedge \omega \in 2^{N \to O} \wedge B \in 2^{V \to VALUES} \wedge D_P(‹\varpi,\omega,B›)\}$

### 4.3.5 Requirement satisfaction

The requirement of a policy is applied against a match to determine whether the policy is violated or not. From an informal point of view, the requirement is satisfied by a match (and the policy not violated) if all the predicates evaluate to true given the match. More precisely, using the variable bindings from the match, a match violates the policy unless each of the node requirement predicates evaluate to true when applied to the object that the corresponding node is associated with in the match and each of the edge requirement predicates evaluate to true when applied to the event that the corresponding edge is associated with in the match.

Formally, in terms of pattern graphs, whenever the domain pattern graph matches, we check to see that the requirement pattern graph also matches. As opposed to the finding a domain match, when matching the requirement pattern graph, the variable bindings and the mappings from the pattern graph to the system graph has already been determined and are represented in the match. Formally noted, $R_P$ is whether the requirement of policy P (as defined in Definition 10) is satisfied by a match ‹$\varpi,\omega,B$›.

**Definition 22.** Requirement satisfaction.
   $R_P(‹\varpi,\omega,B›)$ = match_graph$(G,\varpi,\omega,\lambda,B)$

### 4.3.6 Policy applicability and violation

A policy is *upheld* by a system if and only if all matches of the domain to the system have their requirement satisfied. A policy is *violated* by a system if there is a match where the requirement is not satisfied, which is logical negation of a policy being upheld on the system.

A set of policies are often in effect on a system. The policies are composed to form an overall policy[7] for the system. The overall policy is violated if and only if any of the individual policies are violated. The composed policy is upheld if each of the individual policies are upheld.

We formally define this here. Upheld(P,S) denotes whether a policy or set of policies, P, is upheld on a system S, and violation(P,S) is whether there is a policy violation.

**Definition 23.** Policy semantics for a system.
   For a policy graph P applied to a system S:
      upheld(P,S) = ‹$\forall \Theta: \Theta \in$ matches(P,S) : $R_P(\Theta)$›
      violation(P,S) = ¬upheld(P,S) = ‹$\exists \Theta: \Theta \in$ matches(P,S) : ¬$R_P(\Theta)$›

**Definition 24.** Semantics of policy composition.
   For a set of policy graphs P applied to a system S:

   violation(P,S) = ‹$\exists p : p \in P :$ violation(p,S)›
   upheld(P,S) = ¬violation(P,S) = ‹$\forall p : p \in P :$ upheld(p,S)›

## 5.0 Examples

This section presents several examples of using LaSCO to describe policies. With the aim of exploring the different kinds of policies that LaSCO can state, we loosely segregate all LaSCO policies into a taxonomy, mostly based on the graph syntactic components present. We do not claim that this is the best classification, but we find it convenient. Our top level categories of policies are based on whether edges and isolated nodes are present and how many there are. Section 5.1 discusses policies that have a single edge and Section 5.2 those that have multiple edges. While those policies do not contain isolated nodes, Section 5.3 discusses those that do.

---

7. The composed policy is not a single LaSCO policy graph; rather it is the set of LaSCO policies that it was composed from.

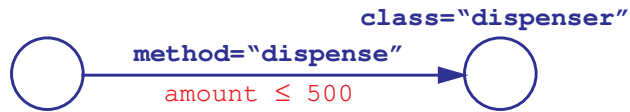
**Figure 7.** Policy graph for the ATM example property

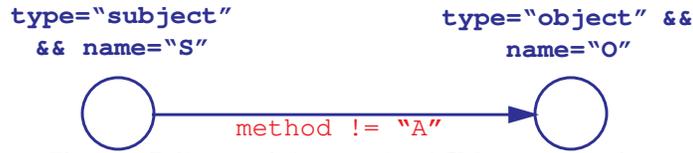
**Figure 8.** Policy graph for negative ACM entry example

## 5.1 Single-event restriction policies

The policies in this section are policies involving just a single event and its associated objects. This corresponds to a LaSCO policy graph containing a single edge and adjacent nodes. Policies like this do not make use of historical context (except inasmuch as the object state records it) and have particularly low complexity for checking whether they are violated or not.

### 5.1.1 State restriction for an access policies

For state restriction for an access policies, the policy relates to a particular type of access, and restricts the state of subjects, objects, or both, or some condition between them. As this type of policy is represented in LaSCO, the type of access that is restricted is represented in the domain predicates. The restriction that is on the state of objects is denoted in the node requirement predicate. Variables in these predicates may be used to state a condition between objects, and if the requirement predicate does not refer to attribute names, the requirement predicate may actually appear on the edge. The access restrictions of Bell-LaPadula [2] and Biba [3] fall into this category. As an example, refer back to the simple security property depiction from Section 2.2, Figure 2.

### 5.1.2 Restricted event parameter access policies

Restricted event parameter access policies limit the values of parameters on a event. This is denoted in LaSCO by an edge requirement predicate that states the restriction on event parameters. An edge domain predicate describes the kind of event being restricted. The policy can additionally be limited in its application to just a certain type of source or destination object. This is achieved through domain predicates on the edge and nodes.

As an example, consider a bank that operates an automated teller machine (ATM). One of their security interests is making sure that the machine does not give out too much money, either through misuse or a flaw in the control system. The ATM control system is object-oriented, and includes a class "dispenser" which interfaces the cash dispenser hardware to the rest of the program. The dispenser class contains a "dispense" method, which tells the dispenser to release a certain amount of money. One policy the bank may wish to impose on the system is that no part of the control system should call the dispense method with an amount parameter greater than 500. This policy is shown in Figure 7.

### 5.1.3 Attribute-based ACM-type entries policies

Policies of this type are restrictions for subjects regarding their access to objects. This is the type of restriction found in access control matrices (ACM) [12] and access control lists (see [6]). Individual restrictions are of a positive tense, indicating that that type of access is allowed or a negative tense, stating that the access is not allowed.

Note that a particular positive subject-object-access ACM entry cannot be represented in LaSCO. This is simply because it does not state a constraint but is merely a piece of a policy. It merely states that something is authorized, without saying, for example, that nothing else is authorized. Negative entries can be expressed though, as shown in Figure 8. This policy is that subject S cannot access object O using access method A. Note for negative ACM entries, the requirement can be imposed on either the subject, object, or the access as a negation of the stated item. For example, Figure 9 depicts an equivalent restriction.

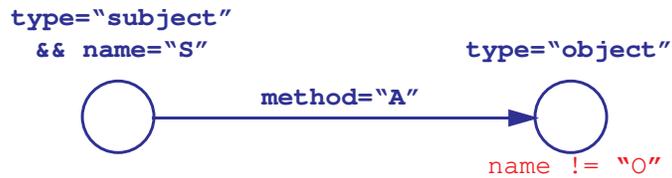

**Figure 9.** Alternate policy graph for negative ACM entry example

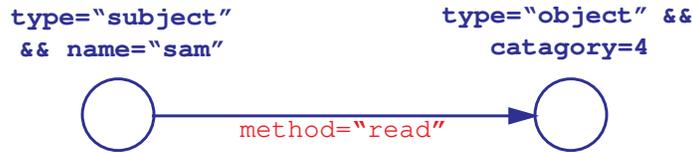

**Figure 10.** Policy graph for attribute ACL example

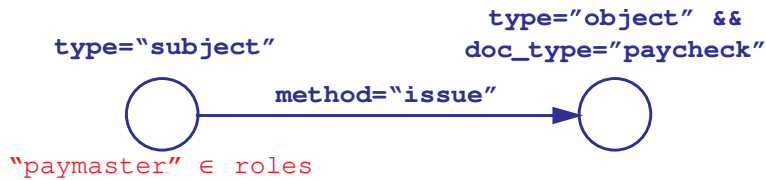

**Figure 11.** Policy graph for payroll RBAC example

Multiple individual ACM triples can be merged with one dimension in common to form a single policy graph. For example, all the valid accesses for a subject and object combination can be considered together and formed into a policy graph. This can likewise be done for all subjects for a particular access type to a particular object and for all objects that can be accessed in a particular way by a particular subject. Here positive authorizations are expressible since we know all the legal (or all the illegal) possibilities in one dimension of the matrix (i.e. over all subjects).

The case of a wildcard access control list entry that makes a policy applicable to all subjects, objects, or accesses is permitted in LaSCO by having that policy entity have no domain predicate. A domain predicate that refers to local attributes or event parameters limits the policy's applicability to objects or events that meet that restriction. For example, Figure 10 depicts the restriction that if subject "sam" is accessing an object in *category* 4, then it must be a read. No other types of access is allowed in this case.

### 5.1.4 Role-based access control policies

Role-based access control (RBAC) (see a review in Sandhu, *et.al.* [15]) is similar to access-matrix type restrictions discussed in Section 5.1.3. The major difference is in the subject. Whereas the subject in an ACM is a user, the subject in RBAC is one of a defined set of roles. Every user with a certain role is treated the same with respect to access control. One might also select objects by attributes.

RBAC is used on a system with defined roles and roles that are (possibly dynamically) assigned to users. One can denote the roles a user currently has active by a *roles* set attribute on objects that are of the type subject. This is part of the way the system is modeled. Figure 11 denotes the RBAC policy that only subjects that have the role of paymaster, can issue a object of document type paycheck.

### 5.2 Multi-event pattern of access policies

For multi-event pattern of access type policies, we need to check a requirement when we have multiple events seen together in a certain pattern. This might be considered a signature of access to find. LaSCO policies for this have more than one edge. We can restrict what each object or event in the signature matches through the domain predicates on its policy node or edge. A couple of subclasses of this are presented in the following subsections.

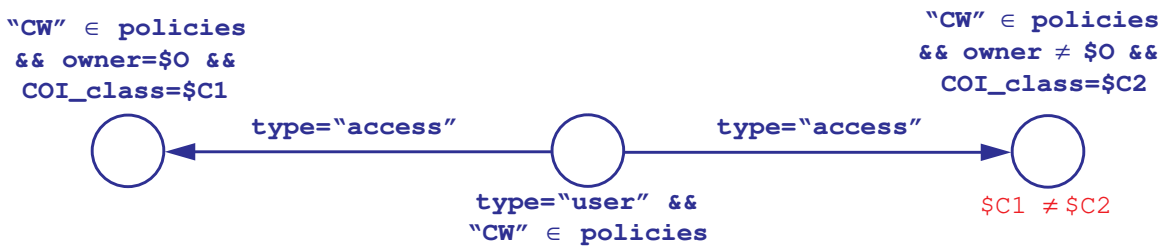
**Figure 12.** Policy graph for the Chinese Wall policy

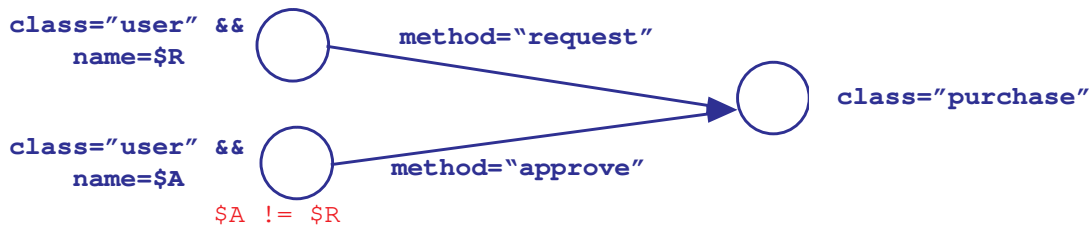
**Figure 13.** Policy graph for purchase request and approval separation of duty

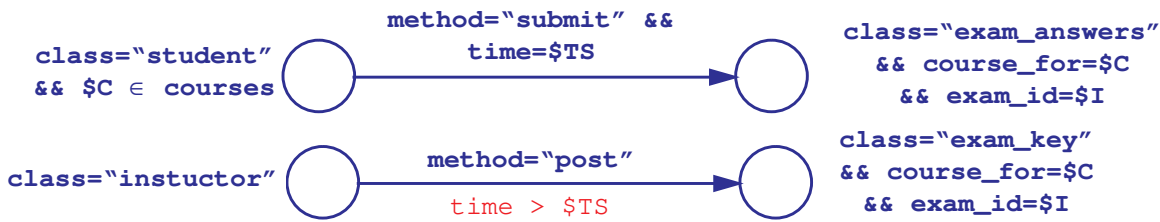
**Figure 14.** Policy graph for ordering of exam submission and key posting restriction

#### 5.2.1 Access comparison policies

For access comparison policies, the policy contains two or more accesses which we compare the details of to determine if the policy is upheld or violated.

Some policies such as Chinese Wall [4] impose restrictions when events that access objects have a common origin. The idea behind the Chinese Wall policy is to prevent conflict of interest situations by consultants that may be employed by a number of parties with competing interests. The policy achieves this by forbidding someone from accessing data from different parties where the parties are in the same conflict on interest class. This is depicted in LaSCO by a node with several edges originating from it as shown in Figure 12. The middle node is a "consultant" whose accesses are limited by the Chinese wall policy. The edges from the consultant node represent accesses to sensitive objects with different owners that are subject to Chinese wall. The constraint is that the owners of these objects cannot be in the same conflict of interest class, stored in the attribute "COI_class".

Separation of duty policies can be depicted in LaSCO and fall under the comparative access policy category. This policy requires that for a particular pair of related accesses, the users involved must be different. This is to avoid conflict of interest problems. In LaSCO this is depicted by two edges with different sources leading to either a single node or to nodes that are linked through their domain predicates. An example is shown in Figure 13. This depicts a policy for a system where there is a separate function for requesting policies and getting them approved. The policy states the restriction that the user that does the "request" must be different than the one that does the "approve".

Time-ordered access restrictions are another type of access comparison policies. What is compared for this type policy is the times of the different events. To do this in LaSCO, we use the *time* parameter on the events and policy variables set to time values for comparison in edge requirement predicates. Figure 13 shows an example of this. This is a policy that might be in effect over a system for electronically processing exams. The restriction is that when a student in a course submits his or her solutions for an exam in the course, then that should take place before the time that an instructor posts the exam key. Here we use the variable C to represent the course and variable I to represent the exam identifier. Variable TS records the time of the submit from the domain for use in comparing it to the time of the post in the requirement.

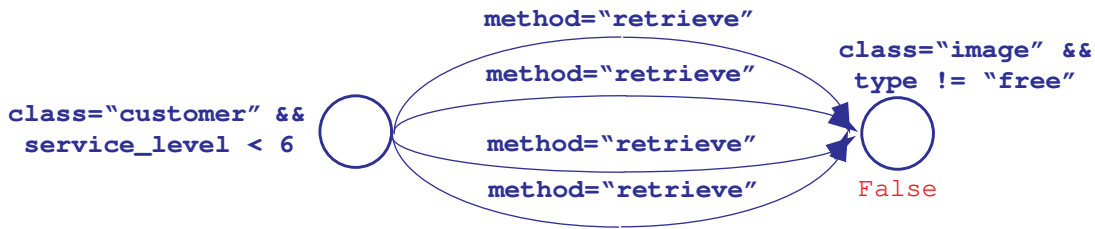

**Figure 15.** Policy graph for image retrieval quantity restriction

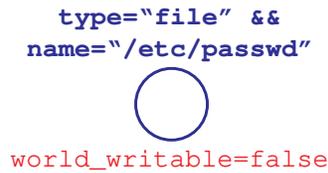

**Figure 16.** Policy graph for password file restriction

### 5.2.2  Count of object accesses policies

LaSCO can be used to impose a maximum number of accesses of a particular type that are allowed. For this, we have the domain describe the situation of the k+1st access where k is the maximum number. Obviously, this is a violation of policy without needing to check anything additional, so we have "False" as an (arbitrarily placed) requirement predicate[8]. Note also that the policy violation could not have been occurred before the domain is satisfied, so this type of policy statement matches at exactly the right instance.

The edges found here represent the type of access that is limited and typically have identical domain predicates. The edges might have a common source node and a common destination node, but this is not always the case. In cases where an object is common to the access edges, the domain predicate on the corresponding policy node restricts the matching node to a certain kind of object. Moreover, if attributes referred to in this predicate may change, then the policy will only count accesses that occur while the object has the attribute with the indicated values.

As an example of this, consider the LaSCO policy depiction in Figure 15, where we show a policy that might be relevant to a program that accesses images from a database. The restriction is that, for a customer with assigned service level less than 6 and for images that are not free, the image cannot be retrieved by the customer more than three times.

## 5.3  Object state policies

LaSCO may be used to state restrictions on objects that do not necessarily have relevant events incident to them. These take the form of policy graphs that contain one or more isolated nodes. The domain for such a node might match an object at any instant and the requirement holds for all such instances. We expect isolated nodes to be used primarily as either an additional domain condition or in a policy graph with just isolated nodes. We present an example of this in Figure 15. The policy in this figure states that the "/etc/passwd" file should never be world writable.

# 6.0  Policy Operations

We sometimes consider operations on LaSCO policies. The result of these operations are new policies which may, or may not, be expressible as a policy graph. Towards the end of operating on policies, we will consider the definition of upheld from Definition 23 to be the statement of what a policy represents in terms of constraints on the system.

---

8. With a "False" predicate, there is never a match.

## 6.1 Policy nullification

The nullification of a policy produces a policy without any effect. Thus the policy is always satisfied. This might be used to disable a policy.

**Definition 25.** Policy nullification.
For a policy P:
$\varnothing P$= true is the nullification of a policy

This may be represented in LaSCO as a pattern graph with arbitrary domain and with requirement predicates that are all "true."

## 6.2 Policy conjunction

Two policies being conjuncted means that they are both in effect. For a given part of a system, a particular policy is relevant if its domain matches and is not relevant if the domain does not match. Thus, none, some, or all of the requirements may need to be satisfied for both policies to be in effect. The result of their satisfaction determines whether the conjuncted policy is upheld or violated. This is the logical "and" of the individual policies.

**Definition 26.** Policy conjunction.
For policies P1 and P2 and a given system S:
$P1 \wedge P2$= ‹$\forall \Theta: \Theta \in$ matches(P1,S) : $R_{P1}(\Theta)$› $\wedge$ ‹$\forall \Theta: \Theta \in$ matches(P2,S) : $R_{P2}(\Theta)$› is the conjunction of P1 and P2

We can similarly define conjunction for more than two policies.

When two conjuncted LaSCO policies have identical domains, the result can be represented as a single LaSCO policy by taking the logical "and" of the requirement predicates for respective nodes and edges. We denote that formally in Definition 27:

**Definition 27.** Policy conjunction for identical domains (graph definition).
‹‹N,E›,$\gamma$,$\lambda_1$,V› $\wedge$ ‹‹N,E›,$\gamma$,$\lambda_2$,V›= ‹‹N,E›,$\gamma$,$\lambda_1 \wedge \lambda_2$,V› where

‹G,$\gamma$,$\lambda$,V› is a tuple denoting a LaSCO policy as in Definition 10,
I=N $\cup$ E,
$\lambda_1 = \cup_{i \in I} \{i \Rightarrow \lambda_1(i)\}$,
$\lambda_2 = \cup_{i \in I} \{i \Rightarrow \lambda_2(i)\}$, and
$\lambda_1 \wedge \lambda_2$ denotes $\cup_{i \in I} \{i \Rightarrow (\lambda_1(i) \wedge \lambda_2(i))\}$

This definition can be used to combine a series of conjuncted policies with identical domains into a single policy graph.

## 6.3 Policy disjunction

Policies are disjuncted when, for cases where their domains overlap, only one of the requirements need be met. In other cases, the policies must be individually met.

**Definition 28.** Policy disjunction.
For policies P1 and P2 and a given system S:
$P1 \vee P2$= ‹$\forall \Theta: \Theta \in$ matches(P1,S) $\wedge$ $\Theta \in$ matches(P2,S) : $R_{P1}(\Theta) \vee R_{P2}(\Theta)$›
  $\wedge$ ‹$\forall \Theta: \Theta \in$ matches(P1,S) $\wedge$ $\Theta \notin$ matches(P2,S) : $R_{P1}(\Theta)$›
  $\wedge$ ‹$\forall \Theta: \Theta \notin$ matches(P1,S) $\wedge$ $\Theta \in$ matches(P2,S) : $R_{P2}(\Theta)$› is the disjunction of P1 and P2
$\vee_{i \in \{1,2,...,n\}} P_i = P_1 \vee P_2 \vee ... \vee P_n$ denotes disjunction amongst a set of policies $\{P_1, P_2, ..., P_n\}$

Disjuncted LaSCO policies generally cannot be represented as a single LaSCO policy, even if the domains are identical.

## 6.4 Policy requirement reversal

By the reversal of a requirement of a LaSCO policy, we mean that while the domain is the same, the requirement is the logical negation of the original. That is, wherever the policy was once upheld, it is now violated and vise versa. For situations under which the domain does not apply in the original policy, it still does not apply.

**Definition 29.** Policy requirement reversal.
For a LaSCO policy P and a given system S:
$\sim P = \langle \forall \Theta: \Theta \in \text{matches}(P,S) : \neg R_P(\Theta) \rangle$ is the requirement reversal of policy P

This is not generally representable as a single LaSCO policy. However, it can be represented as a disjunction of several LaSCO policies. Each of the individual policies have the same domain as the original, but each has one requirement predicate from the original, but negated, and all the other requirement predicates being "true"[9]. We state this formally in Definition 30.

**Definition 30.** Policy requirement reversal (graph definition).
$\sim P = \vee_{i \in I} P_i$ where
P=⟨⟨N,E⟩,γ,λ,V⟩ as in Definition 10,
I=N ∪ E, and
$P_i = \langle\langle N,E \rangle, \gamma, \{i \Rightarrow \neg \lambda(i)\} \cup \cup_{j \in I\text{-}i}\{j \Rightarrow \text{true}\}, V\rangle$

We define the policy requirement reversal for a set of disjuncted LaSCO policies that have the same domain as the conjunction of the policy requirement reversal of each of the individual policies. The semantics for the case where the domains are not identical is that in matches that satisfy both domains, the negation of each of the policies must hold. In the case that only one domain matches, the negation of the requirement for that policy must hold. We define those formally in Definition 31.

**Definition 31.** Policy requirement reversal for disjuncted LaSCO policies.
For a set of LaSCO policies $\{P_1, P_2, ..., P_n\}$, each with identical domains:
$\sim(P_1 \vee P_2 \vee ... \vee P_n) = \sim P_1 \wedge \sim P_2 \wedge ... \wedge \sim P_n$
For arbitrary LaSCO policies P1 and P2 and a given system S:
$\sim(P1 \vee P2) = \langle \forall \Theta: \Theta \in \text{matches}(P1,S) \wedge \Theta \in \text{matches}(P2,S) : \neg R_{P1}(\Theta) \wedge \neg R_{P2}(\Theta)\rangle$
$\wedge \langle \forall \Theta: \Theta \in \text{matches}(P1,S) \wedge \Theta \notin \text{matches}(P2,S) : \neg R_{P1}(\Theta)\rangle$
$\wedge \langle \forall \Theta: \Theta \notin \text{matches}(P1,S) \wedge \Theta \in \text{matches}(P2,S) : \neg R_{P2}(\Theta)\rangle$

## 6.5 Pattern graph coverage

We refer to the set of matches that a pattern graph implies for all possible systems as the coverage of the pattern graph. A pattern graph $G_1$ has greater coverage than pattern graph $G_2$ if $G_1$ matches whenever $G_2$ does and the set of matches for $G_1$ is a larger than the set of matches for $G_2$; has lesser coverage if $G_2$ matches whenever $G_1$ does and the set of matches is smaller; and has equal coverage if the matches occur in exactly the same places.

**Definition 32.** Pattern graph coverage.
For pattern graphs $G_1$ and $G_2$:
$G_1 \supset G_2$ denotes that $G_1$ has greater coverage than $G_2$.
$G_1 \supseteq G_2$ denotes that $G_1$ has greater or equal coverage than $G_2$.
$G_1 \subset G_2$ denotes that $G_1$ has lesser coverage than $G_2$.
$G_1 \subseteq G_2$ denotes that $G_1$ has lesser or equal coverage than $G_2$.
$G_1 = G_2$ denotes that $G_1$ has equal coverage to $G_2$.

## 6.6 Policy containment

A LaSCO policy contains another if the second policy is enforced by the first policy. That is, the second policy can be disregarded if the first is in effect. This occurs if the domain of the enclosing policy has greater or equal coverage them the domain of the enclosed policy and the requirement of the enclosing policy has lesser or equal coverage than the requirement of the enclosed policy (thus it is more restrictive).

---

9. Note that the set of LaSCO policies in the disjunction is finite since there is a finite number of requirement predicates on the original policy graph. Also note that requirement predicates that are "true" need not have a corresponding policy graph created for them as the requirement predicate would be "false". This means that that policy would never contribute to satisfying the disjunction.

**Definition 33.** Policy containment.
For policies $P_1 = \langle G_1, \gamma_1, \lambda_1, V_1 \rangle$ and $P_2 = \langle G_2, \gamma_2, \lambda_2, V_2 \rangle$ as defined in Definition 10:
$P_1 \Rightarrow P_2$ iff $\langle G_1, \gamma_1, V_1 \rangle \supseteq \langle G_2, \gamma_2, V_2 \rangle$ and $\langle G_1, \lambda_1, V_1 \rangle \subseteq \langle G_2, \lambda_2, V_2 \rangle$
$P_2 \Leftarrow P_1$ iff $P_1 \Rightarrow P_2$

## 6.7 Policy operation properties

We believe the following are valid properties of the policy operations we defined above, but have not yet formally proved so:

$\sim(\sim P) = P.$
$P_1 \wedge P_2 = \sim(\sim P_1 \vee \sim P_2)$
$P_1 \vee P_2 = \sim(\sim P_1 \wedge \sim P_2)$

It seems that additional notation for policies will facilitate these proofs.

## 7.0 Discussion

Though aimed at specifying security policies for access control, LaSCO has a more general nature than that. It can specify general system constraint policies which are policies that impose a condition on a system that determines if the policy is upheld or violated. This condition may be based the current state of the system and other various events that have occurred in the history of the system. Example constraint policies are access control, limiting the actions of users, and system assertional checking.

The system model employed by LaSCO is simple, very flexible, and is adequate for modeling the aspects of systems. Traditionally the entities for security have been considered to be subjects and objects. Both of these can be modeled as objects for our system model. These objects interact through accesses and communication, both of which can be modeled as events. However when the system the policy is applied to more closely resembles looking at the state of the system without reference to events, the representation is somewhat awkward. For example some object state that naturally involves two objects (i.e., a subject and an object in an access permission) is represented as one or more object attributes. This results in the policies specified for the system being less conducive to understanding through visual means than policies for other systems. Fortunately this does not seem to be the case frequently since the system could have been modeled more naturally.

Single-edge LaSCO policies permit the specification of simple access control policies where all the information needed to make the decision is available in the current system state and events. LaSCO policies that make reference to multiple events permit access decisions to be based on what has previously occurred. This allows various policies to be specified as shown in the previous section. When isolated nodes appear in a policy, they can impose restrictions of the state of a object without reference to any event. As a result of the flexibility of LaSCO, an application developer or site can create custom policies to fit their needs. This ability promotes security and is in contrast to some policy mechanisms which only allows a limited number of policies to be enforced.

Some constraint policies cannot be stated in LaSCO though. This includes policies that impose requirements to the effect that certain events are required to occur. This is the case for the policy that employees must execute orders given by their supervisor and is also the case when we require a minimum number of events to occur. We discuss an extension to LaSCO to handle this in Section 7.1.4.

Another type of constraint policy that cannot be stated results from two aspects of LaSCO's semantics. One is that the domain predicate on a node must be satisfied with the same variable bindings for all events incident the node. The other is that objects can match at most one node in a policy. This means that we cannot state policies that refer to an object having to be in two or more distinct states through the system. We have not seen a natural instance of this sort of policy though.

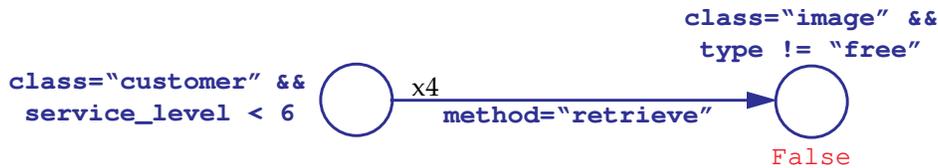

**Figure 17.** Policy graph for image retrieval quantity restriction with edge count extension

In LaSCO, it is not possible to express the fact that in order for the policy to be applicable certain events should have not occurred. An example of this is policies that refer to events that occur without other events having occurred. Section 7.1.7 considers a language extension to allow this.

While LaSCO policies state what a policy violation is, they do not state what to do when a policy violation occurs. This might be an important part of enforcing the policy. An extension to LaSCO towards that end is discussed in Section 7.1.5.

## 7.1 Possible language extensions

This section presents some ideas regarding possible extensions to the LaSCO policy language. These extensions would allow more types of policies to be specified in the language or allow constraints to be stated more naturally. However, any of these extensions would make the language more complicated, which may not be desirable.

### 7.1.1 Edge count

Sometimes, we have a number of identical edges between two nodes in the policy, such as was the case in Figure 15. An extension that would often be convenient in that case is to compress all the edges down into a single edge with a count label on it. This might be depicted as in Figure 17. Edges that are not duplicated can also appear.

### 7.1.2 Additional predicate operations

Beyond the operations defined for the predicates in Section 4.2, we can add more operations that can be used in predicates, allowing for more natural statements of policy. For example, we could allow a regular expression test operation or allow array indexing. It does not seem that adding more operators adds to the number of policies that may be specified, as the system could be modeled to work with the operators that are available. We should be cautious when adding operators as they might complicate the extraction of variable bindings from condition expressions by the extract_bound() function (see Section 4.3.1.3).

### 7.1.3 Global system state

It would be convenient in some situations to model the system as containing some state that is separate from any particular object or event. For example, we might want to be able to refer to the time of day clock, a global security alert state, or a similar security relevant status in the policies. These could then be referred to in node and edge predicates.

A degree of caution is warranted here. With the current language definition, partial matches of the policy to the system can be incrementally kept. This ability might be lost with the global system state extension, resulting in slower checking of policies at run-time.

### 7.1.4 Requirement nodes and edges

Consider an extension to LaSCO which allows pattern nodes and pattern edges to be strictly a part of the requirement portion of the policy. This would allow policies that, for a certain domain, require events to occur to be captured in LaSCO. This means that the basic graph in the pattern graph of the requirement could be different than that in the domain pattern graph. Naturally, this extension would complicate the semantics of the language, opening the question of when (or how quickly) must the required event occur. With current semantics, the requirement must be met when the domain is satisfied. This extension would also decrease the intuitiveness of policy depictions.

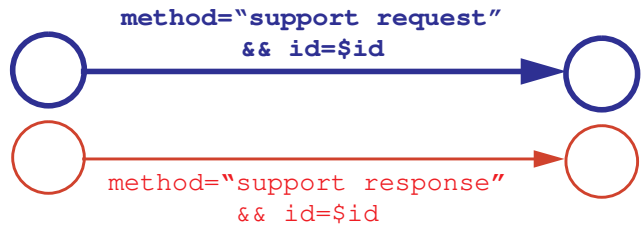

**Figure 18.** Policy graph for support response policy with requirement nodes and edges extension

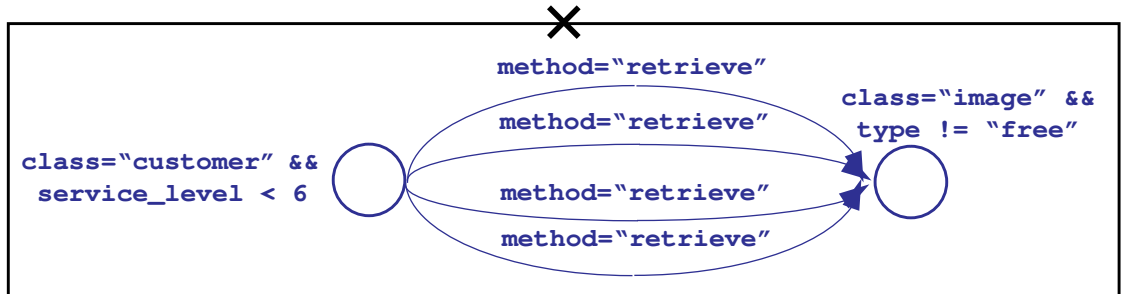

**Figure 20.** Policy graph for image retrieval quantity restriction using the negative policy extension

An example of a policy that this extension would allow to be specified is that all requests for support must be responded to. This is depicted with the language extension is Figure 18. If there were more restrictions, i.e., on who responds, then these could be added to the policy graph.

### 7.1.5  Violation response

When a match for the policy does not satisfy the requirement, the policy is violated. In this case, we might want to specify some response measures to take. Besides preventing the violation, we could log the violation or alert a system administrator. If the policy were used in a intrusion detection system with active response capability, the response might involve blocking the operations of the party that violated the policy or raising the security alert state.

We could extend LaSCO to specify the appropriate response. At least two approaches could be taken here. One is to add into LaSCO general primitives that the particular enforcement mechanism in place could carry out. The other is to provide a generic framework for specifying the responses to be taken. The later case would be enforcement mechanism-specific (and hence the policy would be tied to that mechanism), but would keep LaSCO simpler. It seems that the decision regarding responses would likely be made in the context of the enforcement mechanism anyway though. The response primitive could have function call-style syntax, where the parameters are constants and policy variables. There could be a set of these responses along side the policy graph. Note that in general we would not know the semantics of what the response does.

### 7.1.6  Negative policies

A negative policy is one which indicates what a violation is rather than indicating what a case of the policy being upheld is. The meaning of the requirement is reversed while the domain remains that same, as with the requirement reversal operator in Section 6.4. This could be depicted by an "X" on the policy whereas a positive policy would not have the "X". This changes the semantics of a requirement matching.To check a negative policy, one would find all matches for the domain as before, but then check to see if any of the requirement predicates evaluate to true. If this is the case, then the policy is upheld, otherwise it is violated.

We provide an example of this in Figure 20, which presents the policy from Figure 15 using the extension. Note that this negative policy provides an example of what should not happen.

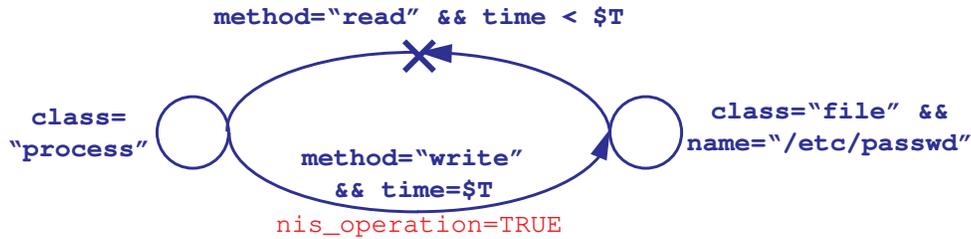

**Figure 21.** Negative edge policy graph for password writing constraint example

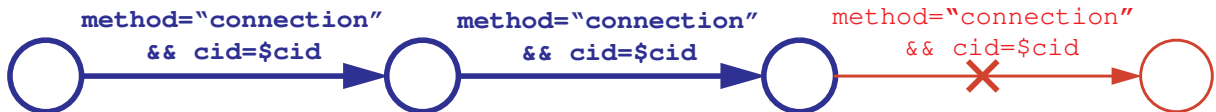

**Figure 22.** Negative consequent edge connection chaining constraint example

#### 7.1.7 Negative edges

Regular policy edges are a pattern for what events should be present in a match of the system. We could extend LaSCO to include negative policy edges, which are a pattern for what events should not have occurred. That is, in order for the domain match to occur, the positive edges should match events that are present and the negative edges should not match any events. It seems that the semantics should be that an event matching a pattern should not have occurred before the rest of the edges match. Otherwise in a runtime policy checker, we could never be sure the policy applies. However, this is an open issue. This extension does not allow additional policies to be specified that are not specifiable in the basic LaSCO language, as the system could be model to present to the policy the knowledge about whether certain events have or have not occurred. This extension, however, might make expressing such policies more natural, and certainly makes it more visual.

An example use of this extension is in shown by the policy graph in Figure 21. The constraint is that if there is a process that does a write to the password file without first reading it, then the write had better be a NIS operation.

If this extension were used in conjunction with the requirement node and edge extension (Section 7.1.4), then a negative requirement-only edge would indicate that the requirement includes a requirement that the indicated type of edge should not be present if the domain matches. As an example, consider the policy depicted in Figure 22. The constraint depicted is that there cannot be a chain of connections more than two deep. The domain in this case is a chain of length 2, and the requirement is that there is no further connections beyond this. The "cid" parameter on edges is an unique identifier for a virtual connections.

## 8.0  Implementation plans

We will implement policy specification and enforcement for Java programs using the architecture described in Section 2.3. From Java source code, we will extract a schema graph. We will develop a schema graph text format to store the extracted schema graph in. This format will likely resemble a Java program. The schema graph extraction mechanism will be based on an existing Java parser.

We have developed a format for representing LaSCO policies and generic LaSCO policies as text and a parser for this format. We have begun implementation of a security policy tool through which LaSCO policies will be specified. This will accept schema graphs, generic policies, and policies as input and will facilitate the development of policies for a schema. This will have graphical user interface for X-Windows and is being built using the Forms Library for X (XForms) [19]. Using the interface, the user can select aspects of the schema with which to instantiate a generic policy and to edit policies.

The policy compiler will accept Java source programs and their schema graphs and the policies that should be enforced by the program. The compiler will insert Java code to enforce the policy. This will take the form of wrappers around certain method invocations and object state updates. These wrappers will note the change in the system and ascertain whether the new action will violate the policy. If it would violate the policy, it can be prevented. The identification of places in the system where a particular policy applies will proceed incrementally as the system executes. Certain parts of the policy can be evaluated statically and will be used to reduce the amount of run-time checking needed. An existing Java parser will be used as a basis for this compiler.

### 9.0  Comparisons with other work

The work of this paper most closely resembles the Miró work of Heydon, Tygar, Wing, *et. al.* at Carnegie Mellon University. Miró consists of two languages, an instance language and a constraint language [10], both of which are based on Harel's hierarchical graphs [9]. The instance language's formal semantics is defined in [13] and is peer to our system graph. It describes a file system access control matrix. The constraint language, which describes security constraints on the file system, contains a domain part (called an antecedent) and a requirement part (called a consequent) and has a predicate on nodes which is similar to LaSCO's predicates. The constraint language for which a constraint checker was implemented for the language for Unix [11]. Unlike Miró, LaSCO states policies in terms of system execution (rather than a snapshot), can be applied to systems other than file systems, and has simpler semantics.

Access matrices [12] and the related access control lists and capabilities (see [6]) are a traditional means of specifying security permissions. While these are in a convenient form for making automated decisions regarding whether an access is authorized or not, their granularity is individual subjects and objects, which does not make them suitable for stating a policy that is less specific in where the policy applies. TAM [16] introduces safety properties into an access control matrix through the use of object typing and defined sets of operations to execute under different conditions. As we have shown in this paper, LaSCO can include historical context in its access control, and its policies can refer to an object of a particular type based its attributes and how the object relates to others in the policy. BEE [14] is an access control mechanism where decisions are based on the result of a boolean expression evaluation for an access right. The goal of this is to allow users to think at a policy level when implementing restrictions and to allow more types of restrictions to be implemented than ordinary access control lists or capability lists. It succeeds to an extent with both of these, though LaSCO has broader means for achieving this. When making decisions, BEE cannot make reference to other events that have occurred. The Authorization Specification Language is a similar approach [8] that expresses the desired authorizations in logic and has conflict resolution rules defined in the language. The goal with this is to enforce varying security policies without changing the security server. This more structured approach towards overall site policies than LaSCO currently defines is useful in certain situations.

Cholvy and Cuppens [5] express the policies on a site in terms of deontic logic, which states what is obliged to occur, what is permitted to occur, and what is forbidden to occur and how to deal with inconsistencies. This seems to be a more general approach than in LaSCO (in which what is obliged is found in the requirement). However, the approach is limited to expressing policies for agents in terms of what they the obliged, permitted, and forbidden from doing. An approach towards specifying authorizations for subjects over objects that is rather formal is presented in Woo and Lam [18]. Precise semantics, allowing for inconsistency and incompleteness in specification, and distributed specification for their logical language are present there. However for this, it is not as clear how to implement the policies for, for example, an application and the patterns for when the policy matches do not seem to be as intuitive.

The Adage architecture [17], developed at the OSF Research Institute, focuses on creating and deploying security policies in a distributed environment. It also focuses on enabling the user to build policy from pieces that the user understands. Their focus is on usability by human users, arguing that security products that user can not understand will not be used. We believe LaSCO can express any Adage policy, but offers the additional benefits of application to different kinds of systems, direct linking to an application program, and formal semantics.

Deeds, developed by Edjlali, Acharya, and Chaudhary [7], is a history-based access control mechanism for Java whose goal is to mediate accesses to critical resources by mobile code. LaSCO can also be used for this purpose. As we plan to do, they insert code into Java programs. However, whereas their basis for access control decisions is the result of dynamically executing Java code provided by the user, our basis is clearly stated policies. We find our approach appealing since it permits conceptual understanding of the access restriction and formal reasoning about the policy.

## 10.0 Conclusion and future work

In this paper, we have presented a formal policy language based on graphs. Using LaSCO, one can specify constraint policies which constrain the use of resources on a system under given circumstances. What is required might depend on the context in which an access occurs. We presented an overview of using LaSCO to specify policies in an object-oriented program. The language was described, its semantics presented, and we discussed what it could and could not specify.

Unlike traditional approaches, LaSCO permits policies to be applied to applications at the language level in addition to at the operating system and runtime system. This has benefits toward specifying policies at the right level of abstraction, toward static policy checking, and toward automatic tool generation. LaSCO has the benefits of allowing policies to be stated separately from the system, of being formally based with formal semantics, and allowing policies to be specified using flexible patterns.

With this policy language, one could construct an enforcement mechanism for a system to implement a set of specified policies. A few example security applications where LaSCO might be used are:

- wrapping or monitoring trusted or untrusted applications to detect activity that violates policy
- an intrusion detection system to monitor a network and report policy violations
- software to monitor or intercept file system accesses to support more expressive access control that is otherwise possible, and
- a monitor of operating system activity looking for undesirable behavior by users and programs.

Though designed for security policies for access control, LaSCO may also be useful for other areas such as software development where behavioral assertions might be checked at run time.

We are implementing LaSCO in Java programs through instrumenting the programs with policy statements checked at run time. In this environment, we can take advantage of restrictions on classes of objects and on the particular event that is involved for the policy. As future work, we will continue our implementation and continue our study of policy operations.

# Security Policy Specification Using a Graphical Approach

James A. Hoagland, Raju Pandey, and Karl N. Levitt

hoagland@cs.ucdavis.edu
UC Davis Computer Science Technical Report CSE-98-3